\documentclass[floatfix,aps,prl,showpacs,amsmath,nofootinbib,
preprintnumbers]{revtex4}

\begin{document}

\preprint{FERMILAB-PUB-04-283-A}

\title{Cosmological influence of super-Hubble perturbations}
\author{Edward W. Kolb}\email{rocky@fnal.gov}
\affiliation{Particle Astrophysics Center, Fermi
       	National Accelerator Laboratory, Batavia, Illinois \ 60510-0500, USA \\
       	and Department of Astronomy and Astrophysics, Enrico Fermi Institute,
       	University of Chicago, Chicago, Illinois \ 60637-1433 USA}

\author{Sabino Matarrese}\email{sabino.matarrese@pd.infn.it}
\affiliation{Dipartimento di Fisica ``G.\ Galilei,'' Universit\`{a} di Padova, 
        and INFN, Sezione di Padova, via Marzolo 8, Padova I-35131, Italy}

\author{Alessio Notari}\email{notari@hep.physics.mcgill.ca}
\affiliation{Physics Department, McGill University, 3600 University Road,
             Montr\'eal, QC, H3A 2T8, Canada\\ and
             Scuola Normale Superiore, Piazza dei Cavalieri 7, Pisa I-56126 
             and INFN, Sezione di Pisa, Italy}

\author{Antonio Riotto}\email{antonio.riotto@pd.infn.it}
\affiliation{INFN, Sezione di Padova, via Marzolo 8, I-35131, Italy}

\date{\today}

\begin{abstract}
\noindent
The existence of cosmological perturbations of wavelength larger than the
Hubble radius is a generic prediction of the inflationary paradigm. We 
provide the derivation beyond perturbation theory of a conserved quantity
which generalizes the linear comoving curvature perturbation. As a by-product,
we show 
that super-Hubble-radius (super-Hubble) perturbations have no physical
influence on local observables (\textit{e.g.,} the local expansion rate) if
cosmological perturbations are of the adiabatic type.
\end{abstract}

\pacs{98.80.cq}

\maketitle


Inflation is an elegant explanation for the flatness, horizon, and monopole
problems of the standard big-bang cosmology \cite{guth81}. But perhaps the most
compelling feature of inflation is a theory for the origin of density
perturbations (the seeds for the large-scale structure of the
Universe) and anisotropies in the cosmic microwave background (CMB)
\cite{lrreview}. Density and gravitational-wave perturbations are created 
during inflation from quantum fluctuations and ``redshifted'' to sizes larger
than the Hubble radius ($R_H\equiv H^{-1}$). They are then ``frozen'' until
sometime after inflation when they once again come within the Hubble
radius. The last and most impressive confirmation of this idea has been
provided by the data of the Wilkinson Microwave Anisotropy Probe \cite{wmap1}.

A general feature of inflation is the existence of scalar perturbations of
wavelength larger than the Hubble radius.  During inflation a small region of
size less than the Hubble radius grows large enough to encompass easily the
comoving volume of the entire presently observable Universe. This requires a
minimum number of \textit{e}-foldings, $N\gtrsim 60$, where $N$ measures the
logarithmic growth of the scale factor during inflation. However, most models
of inflation predict a number of \textit{e}-foldings that is, by far, much
larger than 60 \cite{lrreview}. This amounts to saying that there is a huge
phase space for super-Hubble perturbations.  These super-Hubble perturbations
will re-cross the Hubble radius only in the very far future.

This paper deals with the issue of the physical influence of the infrared
modes produced during inflation on cosmological observables such as the local
Hubble expansion rate. The question is, can we see beyond the horizon and learn
the nature of the very long-wavelength perturbations. This question has already
been posed in the literature with contradictory answers
\cite{mab,u,gb,fmvv,gb2,bl}. In this note we demonstrate that 
if adiabaticity holds, 
super-Hubble perturbations do not  have an impact on local
physical observables, for instance, the local Hubble expansion rate.
Our results are valid at any order in perturbation theory.

Let us first review some generalities which will turn out to be useful.  Since
we are interested in very long-wavelength perturbations, from now on we will
neglect spatial gradients. Our starting point is the Arnowitt-Deser-Misner
(ADM) formalism, with metric
\begin{equation}
ds^2=-N^2\,dt^2+N_i\,dt\,dx^i+\gamma_{ij}\,dx^i\,dx^j . \nonumber
\end{equation}
The three-metric $\gamma_{ij}$, and the lapse and the shift functions $N$ and
$N_i$, describe the evolution of the timelike hypersurfaces. The extrinsic
curvature three-tensor is
\begin{equation}
K_{ij}=\frac{1}{2N}\left(N_{i|j}+N_{j|i}-
\frac{\partial \gamma_{ij}}{\partial t}\right)  , \nonumber
\end{equation}
where three-space covariant derivatives with connection coefficients determined
from $\gamma_{ij}$ are indicated by a vertical bar. The trace $K=K^i_{\ i}$ is
the generalization of the Hubble parameter of homogeneous isotropic
cosmologies. The traceless part of the tensor is denoted by an overbar,
$\overline{K}_{ij}=K_{ij}-\frac{1}{3}\,K\,\gamma_{ij}$. In the ADM formalism
the equations simplify considerably if we set $N^i=0$.

It is convenient to express the spatial metric as \cite{sb}
\begin{equation}
\gamma_{ij}=\exp\left[2\alpha(t,x^i)\right] \, h_{ij}(x^i) , \nonumber
\end{equation} 
where the conformal factor $\exp[\alpha(t,x^i)]$ may be interpreted as the
spatial-dependent scale factor. The time-independent three metric $h_{ij}(x^i)$
(with determinant equal to unity) describes the three geometry of the
conformally transformed space. Within linear perturbation theory,
$\alpha(t,x^i)$ would contribute only to scalar perturbations, whereas $h_{ij}$
contains another scalar, as well as vector and tensor perturbations. Vector and
tensor perturbations are necessary to satisfy Einstein equations, however in
the long wavelength limit they do not affect the equations for the scalar
perturbations.  On super-Hubble scales, $(\partial\overline{K}^i_{\ j}/\partial
t)=N K \overline{K}^i_{\ j}$ and $K=-(3/2 N)\dot{\alpha}$. This implies that
$\overline{K}^i_{\ j}$ decays with time and can therefore be safely set to
zero. Since $\exp[\alpha(t,x^i)]$ is interpreted as a scale factor, we can use
the Hubble parameter $H(t,x^i)=\dot{\alpha}(t,x^i)/N(t,x^i)\equiv -K(t,x^i)/3$
in place of the trace $K$ of the extrinsic curvature.

Now consider a Universe filled with some fluid(s) described by an
energy-momentum tensor of the form $T_{\mu\nu}=
\left(\rho+P\right)\,u_\mu u_\nu +P\,g_{\mu\nu}$, where $\rho$ and
$P$ are the energy  density and pressure, respectively. The four-velocity
vector can be chosen to be $u^\mu=(1,\vec{0})$. In the case of one fluid, this
amounts to saying that a volume measured by a local observer is comoving with
the energy flow of the fluid. In the multi-fluid case, the
volume comoving with the total energy density $\rho$ is not the same as the
individual volumes comoving with the energy density components. However, one
can show that on super-Hubble scales all comoving volumes become equivalent
\cite{lw}, and therefore our choice of the four velocity is well justified.

The relevant set of equations is provided by the continuity equation,
$N^{-1}\partial\rho/\partial t=-3H(\rho+P)$, and by $N^{-1}\dot{\alpha}=H$.
Eliminating the lapse function $N$ from the set of equations, we derive
\begin{equation}
\dot\alpha +\frac{1}{3}\frac{\dot\rho}{\left(\rho+P\right)}=0  .
\label{b}
\end{equation}
The quantity 
\begin{equation}
{\cal F} \equiv \alpha +\frac{1}{3}\int^\rho\, 
\frac{d\rho'}{\left(\rho'+P'\right)}
\label{fun}
\end{equation}
is therefore conserved in time at any order in perturbation theory. To
understand its physical significance, consider an adiabatic fluid for which the
pressure is a unique function of the energy density $P=P(\rho)$. If we set
\begin{equation}
\label{psi}
\exp\left[\alpha(t,x^i)\right]=a(t)\,\exp\left[-\psi(t,x^i)\right] 
\end{equation}
and expand Eq.\ (\ref{fun}) to first order in perturbation theory, we find that
the perturbation $\delta_1{\cal F}$ coincides with the comoving curvature
perturbation
\begin{equation}
\zeta_1=-\psi_1 -H \frac{\delta\rho_1}{\dot{\rho}_0}  , \nonumber
\end{equation}
where $\rho_0$ is the background value of the energy density. Similarly,
expanding Eq.\ (\ref{fun}) to second order in perturbation theory, one finds
that the perturbation $\delta_2{\cal F}$ coincides with the definition of the
second-order comoving curvature perturbation $\zeta_2$ \cite{mw} (up to a
constant piece $\zeta_1^2$)
\begin{eqnarray*}
\zeta_2&=&-\psi_2-H\frac{\delta\rho_2}{\dot{\rho}_0}-
2\frac{H}{{\dot{\rho}_0}^2}
\delta\dot\rho_1\delta\rho_1-2\frac{\delta\rho_1}{\dot{\rho}_0}\left(
\dot{\psi}_1+2H\psi_1\right) \\
& & + 
\left(\frac{\delta\rho_1}{\dot{\rho}_0}\right)^2\left(H\frac{\ddot{\rho}_0}{
\dot{\rho}_0}- \dot H-2H^2\right)  .
\end{eqnarray*}
We conclude that $\delta {\cal F}$ is the nonlinear generalization of the
comoving curvature perturbation on super-Hubble scales for adiabatic
fluids. Equivalently, $\psi$ plays the role of the nonlinear generalization of
the comoving linear gravitational potential.

For a generic nonadiabatic fluid, the perturbation of the pressure energy
density $\delta P$ can not be expressed in terms of the perturbation of the
energy density $\delta\rho$. Perturbing Eq.\ (\ref{fun}) to first order, one
recovers the nonconservation of the comoving curvature perturbation
\begin{equation}
\dot{\zeta}_1=-\frac{H}{\left(\rho+P\right)}\,\delta P^\mathrm{nad}_1 ,
\nonumber
\end{equation}
where $\delta P^\mathrm{nad}_1=\delta P_1-(\dot P/\dot\rho)\,\delta\rho_1$ 
is the first-order nonadiabatic pressure perturbation.

Let us now return to the physical influence of super-Hubble modes. We will
consider the effect on the local expansion rate of the Universe (although an
effect on the local expansion rate suggests that there will be effects on other
cosmological measurables as well). We choose to work in the synchronous gauge
for which $N=1$ (together with $N^i=0$). In this gauge the field equations look
just like the familiar equations point by point. As mentioned, for a generic
set of fluids we may safely take the four velocity to be $u^\mu=(1,\vec{0})$ on
super-Hubble scales, from which we define the local expansion rate of
tangential surfaces orthogonal to the fluid flow, \textit{i.e.,} the local
expansion rate, to be $\frac{1}{3}D_\mu u^\mu$. In the synchronous gauge this
quantity and $H$ coincide.

Using Eq.\ (\ref{psi}), we find that at any order in perturbation theory the
local expansion rate is $H=\dot{a}/a-\dot\psi$.  We immediately infer that the
local physical expansion rate is influenced by long-wavelength perturbations
only if the gravitational potential is time dependent.\footnote{Recall that we
have ignored spatial gradients in $\psi$. A perturbative calculation of the
effect of spatial gradients on the expansion rate was considered in
\cite{KMNR}.} Let us first analyze the case of an
adiabatic fluid. From Einstein's equations for a globally flat space we deduce
(on super-Hubble scales)
\begin{equation}
\label{j}
3H^2=8\pi G \rho =-8\pi G\,P + 2\dot H  .
\end{equation} 
If the pressure $P$ is a unique function of the energy density, we may expand
$\rho=\rho_0+\Delta\rho$, where $\rho_0$ is the energy density entering the
homogeneous Einstein equations from which the homogeneous scale factor $a(t)$
is computed. Inserting this expansion into Eq.\ (\ref{j}) and eliminating the
pressure in favor of the energy density $\rho\propto H^2$, we find a
differential equation of the generic form ${\cal
G}\left(\ddot\psi,\dot\psi\right)=0 $, which does not contain any term
proportional to $\psi$. Therefore, $\psi=\psi(x^i)$ is a solution of Einstein
equations at any order in perturbation theory and $\dot\psi=0$ holds on
super-Hubble scales.  This result reproduces the familiar result that the
gravitational potential does not depend upon time for super-Hubble modes if
there is only one fluid in the Universe. As a by product of Eq.\ (\ref{fun}),
we see that the perturbation of the energy density vanishes on super-Hubble
scales in the synchronous gauge at any order in perturbation theory.

From this generic argument we conclude that there is no influence of very long
wavelength modes on the physical local expansion rate of the Universe if
adiabaticity holds (again, we have ignored gradients in $\psi$).  In the
adiabatic case, the influence of infrared modes is not locally
measurable. There is a simple explanation of this result. When adiabaticity
holds, the pressure is a well defined function of the energy density. This
means that the Hubble rate is only a function of the unique available physical
clock, the energy density. Indeed, suppose that the local expansion rate is
$H(t,x^i)=H(\rho(t,x^i),t)$.  This leads to
\begin{eqnarray}
\left(\frac{\partial H}{\partial t}\right)_{x^i}&=&
\left(\frac{\partial\rho}{\partial t}\right)_{x^i}
\left(\frac{\partial H}{\partial \rho}\right)_{t} +
\left(\frac{\partial t}{\partial t}\right)_{x^i}
\left(\frac{\partial H}{\partial t}\right)_{\rho}\nonumber\\
&=&-4\pi G(\rho+P) +\left(\frac{\partial H}{\partial t}\right)_{\rho}  ,
\label{mmm}
\end{eqnarray}
where we have made use of Eq.\ (\ref{b}) evaluated in the synchronous gauge,
$3H=-\dot\rho/(\rho+P)$. Comparing Eq.\ (\ref{mmm}) with $\dot H=-4\pi
G(\rho+P)$, we see that $(\partial H/\partial t)_{\rho}=0$, and hence
$H(t,x^i)=H(\rho(t,x^i))$. The dependence of the local expansion rate
of the Universe on the clock time takes the same form as in the unperturbed
Universe when evaluated at a fixed value of the only clock available, the
energy density $\rho$.  Infrared modes do not have any locally measurable
effect on the expansion rate of the Universe. This result applies, in
particular, during inflation if the energy density is dominated by a single
inflaton field.

Our findings suggest that adiabatic super-Hubble modes cannot modify
cosmological observables like the Hubble rate.

\section*{Acknowledgments}
E.W.K.\ is supported in part by NASA grant NAG5-10842 and by the Department of
Energy.



\end{document}